\documentclass[11pt]{article}
\usepackage[dvips]{graphics, color}
\usepackage{latexsym}
\usepackage{amsmath}
\usepackage{amssymb}
\begin{document}

 \title{Detectability of Upgoing Sleptons}

 \author{Chao Zhang\thanks{Email: zhangchao@mail.ihep.ac.cn} and YuQian Ma \\
 \small\it{ Institute of High Energy Physics, Chinese Academy of
Sciences},\\ \small\it{ P.O. Box 918-3, Beijing 100039, People's
Republic of China}}
\date{}
 \maketitle

\begin{center}
\begin{minipage}{130mm}
\vskip -5mm
 {\large \bf Abstract:} \, Some supersymmetric models predicted long lived charged
 sleptons. Following Albuquerque, Burdman and Chacko\cite{albu},
 we consider an exact process where the $ \widetilde{\tau}$,
 the supersymmetry partner of $\tau$ lepton, are produced inside the earth by collisions
 of high energy neutrinos with nucleons. We detailedly investigate the possible signals of
 upgoing $ \widetilde{\tau}$ by comparing with the background muons where atmospheric neutrino flux is taken into account. The realistic spectra shows that
 km-scale
 experiments could see as many as 69 events a year by using the Waxman-Bahcall limit on the extraterrestrial neutrino flux.   
     \\

\end{minipage}
\end{center}

\section{Introduction}
\mbox \\

Substantial evidences exist suggesting that most of the Universe's
matter is non luminous \cite{evident1}\cite{evident2}. There are
many predictions for its composition but the nature of dark matter
is yet unknown. \\

Weak scale supersymmetric theories of physics beyond the standard
model, provide perhaps the most promising candidates for dark
matter\cite{susy}. However, supersymmetry must be broken at the
energy scale accelerator having reached since the superpartners have
not been observed yet. Supersymmetric models typically have a
symmetry, called R-parity, which exclusively ensures that the
Lightest Supersymmetric Particle (LSP) is the most stable. Obviously,
the LSP is the natural candidate for dark matter. Different scales
of supersymmetry breaking determine what the LSP is. Typically, if
supersymmetry is broken at high scales, the LSP is likely to be
the neutralino or the quintessino\cite{bi}, however, if
supersymmetry is broken at lower scales, the LSP is likely to be
the gravitino\cite{gravitino}. In the models where the LSP is
typically the quintessino or the gravitino, the Next to Lightest
Supersymmetric Particle (NLSP) tends to be a charged slepton,
typically the right-handed $ \widetilde{\tau}$, which has a long 
lifetime between
 microsecond and a year around\cite{bi}\cite{gravitino}, depending
on the scale of supersymmetry breaking and the slepton's mass. Of
course, these lifetimes are negligible comparing with the age of
our universe, therefore, almost all these NLSP produced in the
evolutive history of the
universe decayed into LSP. \\

Lest to mislead the reader, we should clarify several points. It
is not absolutely certain that the NLSP is either charged slepton
or other neutral supersymmetric particle; furthermore, there is at
present no direct evidence for the existence of supersymmetry.
These are still unproven ideas. But some theoretical conclusions
 prefer the slepton to other neutral NLSPs \cite{slepton}.
Although speculative, supersymmetric
dark matter is very well motivated and based on a simple physical principle.\\

In the models of Slepton as NLSP, since the slepton is massive,
actually, it can possibly be produced by high energy process.
Collisions of high energy neutrinos with nucleons in the earth at
energies above threshold for supersymmetric production can produce
supersymmetry pairs which are unstable and promptly decay into
slepton; since the slepton is charged and long lived, it is worthy
to see whether its upgoing tracks could be detected by some large
cosmic ray experiments or neutrino telescopes, such as L3+C
\cite{l3c}, ICECUBE \cite{icecube}, Super-K \cite{sk}, etc.
Based on a gravitino-LSP scenario, Albuquerque et al initiated that 
 one can take neutrino telescopes as a direct probe of supersymmetry
breaking\cite{albu}. Motivated by their work, our
following calculations base on the quintessino-LSP scenario from Ref.\cite{bi}
where it restricts the stau mass between $100 GeV$ and $1 TeV$ and
the lifetime between $10^{6}\sim 10^{7}$ seconds.

\section{Fundamental Process Analysis}
\mbox \\

\begin{figure}
 \setlength{\unitlength}{1cm}
\begin{picture}(12,7)
\thicklines \multiput(0,0)(6,0){2}{ \put( 3.2,1.5){$\chi^{0}$}
\put( 3.2, 5){$ \chi^{+} $} \put( 4, 2.52){$ \widetilde{\nu} $}
\put( 4, 6.1){$ \widetilde{l} $}
 \multiput(1,0)(0,3.5){2} {
 \put(0, 2.5){\line(1,0){2}}
\put(1,2.52){$\nu$} \put(0, 0.5){\line(1,0){2}}
\multiput(2,0.5)(0,0.4){5}{\qbezier(0,0)(0.1, 0.1)(0,
0.2)\qbezier(0,0.2)(-0.1, 0.3)(0, 0.4)}
\multiput(2.1,0.5)(0.4,0){5}{\line(1, 0){0.2}}
\multiput(2.1,2.5)(0.4,0){5}{\line(1, 0){0.2}} } }
\put(1.8,0.05){$u(d)$} \put(1.8,3.55){$u(d)$}
\put(7.8,0.05){$\overline{u}(\overline{d})$}
\put(7.8,3.55){$\overline{u}(\overline{d})$}
\put(3.8,0.05){$\widetilde{u}(\widetilde{d})$}
\put(3.8,3.55){$\widetilde{d}(\widetilde{u})$}
\put(9.8,0.0){$\overline{\widetilde{u}}(\overline{\widetilde{d}})$}
\put(9.8,3.5){$\overline{\widetilde{d}}(\overline{\widetilde{u}})$}
\end{picture}
\caption{\label{feynman} Feynman diagrams for supersymmetric
particles production in $\nu N$ collisions. the upper two diagrams
refer to charged current interactions, and the under two diagrams
account for neutral current process, with the exchanges of the
chargino $ \chi^{+} $ and neutralino $ \chi^{0} $, respectively.}
\end{figure}
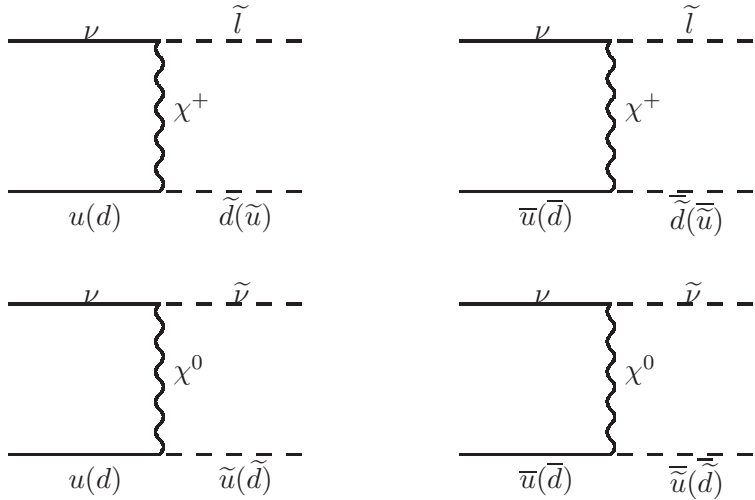

High energy neutrinos interacting with nucleons will produce
supersymmetric particles, and that will promptly decay into a pair
of NLSPs, which have long lifetime. This lifetime is definitely
large enough so that $\widetilde{\tau}$ do not decay inside the
earth.  $\nu N \rightarrow \widetilde{l}\widetilde{q}\rightarrow
2\widetilde{\tau}$, the dominant process is analogous to the
standard model charged current interactions, i.e, $\nu_{\mu} N
\rightarrow \mu^{-}+ anything$. We include the corresponding
processes at the parton
level in Figure \ref{feynman}. \\

\begin{figure}
\scalebox{0.5}{\includegraphics*[0pt,170pt][560pt,670pt]{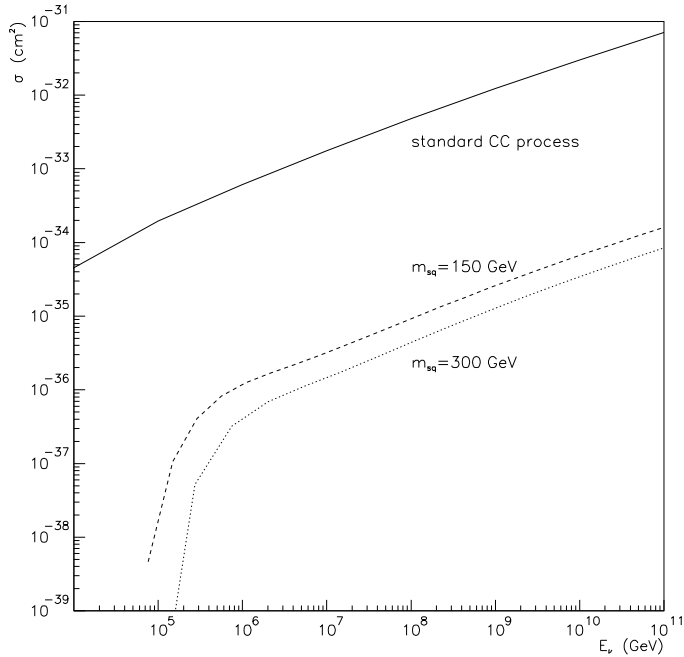}}
\caption{\label{cross} The cross section of $\nu N$ interactions
as a function of neutrino energy. The curves correspond to
$m_{\widetilde{\tau}}=143 GeV$ and for squark masses
$m_{\widetilde{q}}=150, 300 GeV$. The top curve refers to the
standard model charged current interactions. }
\end{figure}

We make use of the stau mass $m_{\widetilde{\tau}}=143 GeV$ and
two values for the squark masses: $m_{\widetilde{q}}=150, 300 GeV$
\cite{bwz} in our calculations, and the cross section for
supersymmetric production as a function of the neutrino energy is
given in Figure \ref{cross}. Also plotted for comparison is the
standard model charged current cross section \cite{cc}. From
Figure \ref{cross} one can conclude supersymmetric interactions is
much weaker than that of standard model.

\section{Calculate upgoing fluxes}
\mbox \\

 In order to be comparable with the results form Ref.\cite{albu}, we 
consider the  similar physical process.
We take the earth as the target and consider the
incoming of some diffuse fluxes of high energy neutrinos. In our
calculations, we make use of a model of the earth density profile
as detailed in
Ref.\cite{cc}.\\

Dynamic analysis shows that the energy threshold for $\widetilde{\tau}$
production has to be over $\sim 100 TeV$. For atmospheric
neutrino flux, usually experimental data give its
 ranges from $1$ to $10TeV$,
which seems out of our need, however, theoretical results
\cite{atmos} show us that atmospheric neutrino spectrum
with energy above $10TeV$ still act an important flux for
$\widetilde{\tau}$ and muon production. We make use of atmospheric
neutrino flux with energy under $10TeV$ in Ref.\cite{zhqq}, and the 
flux with energy above $10TeV$ in Ref.\cite{atmos}.\\
  
\begin{figure}
\scalebox{0.5}{\includegraphics*[0pt,170pt][540pt,670pt]{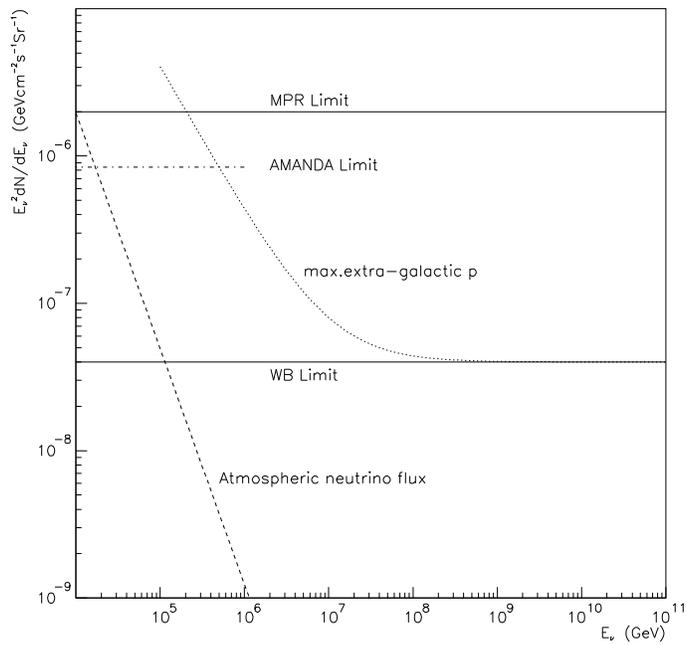}}
\caption{\label{flux} Neutrino fluxes given by different models.
 The  WB bound line gives the upper bound corrected for
neutrino energy loss duo to redshift and for the maximum known
redshift evolution.
 The dot curve is the maximum contribution due
to possible extra-galactic component of lower-energy. The dash-dot
line shows the experimental upper bound on diffuse neutrino flux
established by the AMANDA experiment. The top line is given by
authors  Mannheim, Protheore, and Rachen (MPR) with considering
optically thick AGN models or very strong magnetic fields.
Dashed line is the theoretic predict for 
energetic atmospheric neutrino flux.}
\end{figure}

 Figure
\ref{flux} summarizes some popular theoretical and experimental
upper limits on diffuse neutrino fluxes. The Waxman-Bahcall (WB)
upper bound \cite{wb} assumes that $100\%$ of the energy of cosmic
ray protons are lost to $\pi^{+}$ and $\pi^{-}$ and that the
$\pi^{+}$ all decay to muons that also produce neutrinos.
Ref.\cite{wb} also discussed the maximum contribution due to
possible extra-galactic component of lower-energy $< 10^{17}eV$,
where protons have been first considered (max.extra-galactic p).
Experimentally, AMANDA experiment gave a upper bound on diffuse
neutrino flux \cite{amanda}. By considering optically thick AGN
models or by involving very strong magnetic fields, Mannheim,
Protheore, and Rachen (MPR) have argued that one might be able to
avoid the WB limit and get a higher upper bound \cite{mpr}. From
Figure \ref{flux}, actually, for extraterrestrial neutrino flux,
we can conclude that the upper bound
has been decided by AMANDA experiment on the neutrino energy below
$10^{6}GeV$, therefore, we take ``max.extra-galactic p" as
supplement for the neutrino energy above $10^{6}GeV$. In this
article, unless  extra specification, the following results all
make use of atmospheric neutrion flux with energy above
$1GeV$ adding the conservative extraterrestrial WB Limit as the
 incoming neutrino flux.\\

There are two factors must be taken into account: one is the
attenuation of high energy neutrinos through the earth, the other
is the energy loss of charged $\widetilde{\tau}$
 before running into a detector.\\

The earth is opaque to ultra-high energy neutrinos. When the
energy of neutrinos is over $40 TeV$, the $\nu N$ interaction
length turns out to be larger than the diameter of the earth, the
attenuation of neutrinos must be considered.
\begin{equation}\label{atte}
\mathcal{I}_{\nu}(E_{\nu},x)=\mathcal{I}^{0}_{\nu}(E_{\nu})\exp(-x/l),
\end{equation}
where $\mathcal{I}^{0}_{\nu}(E_{\nu})$ is the initial incoming
neutrino flux, $x$ is the depth a neutrino penetrating the earth,
and $l$ refers to the interaction length, $l\equiv
1/(\sigma_{tot}\cdot n)$, $n$ accounts for the number density of
the medium. Here $\sigma_{tot}$, in principle, include all charged
and neutral current processes contributing by both standard model
and supersymmetric production.  Since the initial interactions
produce slepton and these are nearly degenerated in flavor, the
flavor of the initial neutrino does not affect the results.\\

The upgoing $\widetilde{\tau}$ event rate depends on the
 $\nu N$ cross section
in two ways: through the interaction length which governs the
attenuation of the neutrino flux due to interactions in the earth,
and through the probability that the neutrino converts to a $\widetilde{\tau}$
energetic enough to arrive at the detector with
$E_{\widetilde{\tau}}$ large than the threshold energy
$E_{\widetilde{\tau}}^{min}$. The probability that 
a $\widetilde{\tau}$ produced
in a $\nu N$ interaction arrives in a detector with an energy
above the $\widetilde{\tau}$ energy shreshold $E_{\widetilde{\tau}}^{min}$
depends on the range $R$ of a $\widetilde{\tau}$
 in rock, which follows from the
energy-loss relation \cite{gaisser}
\begin{equation}\label{loss}
-\mathrm{d}E_{\widetilde{\tau}}/\mathrm{d}x=
\alpha+E_{\widetilde{\tau}}/\xi,
\end{equation}
here, the coefficients $\alpha$ and $\xi$ characterize the
ionization and radiation losses respectively. For numerical
estimates of ionization loss here we use $\alpha=2 MeV/(gcm^{-2})$
\cite{ionization}. For a given momentum impulse, the radiation
energy loss is inversely proportional to the square of the mass of
the radiation particle. Thus the radiation length for $\widetilde{\tau}$ is
approximately $(m_{\widetilde{\tau}}/m_{\mu})^{2}$ times large than that
for muons. We take $\xi_{\mu}\approx 2.5 \times 10^{5} g/cm^{2}$
in rock \cite{gaisser} and define $\epsilon \equiv \alpha\xi$,
then the range can be expressed
\begin{equation}\label{range}
R(E_{\widetilde{\tau}},E_{\widetilde{\tau}}^{min})=
\xi\ln\frac{\epsilon+E_{\widetilde{\tau}}}{\epsilon+E_{\widetilde{\tau}}^{min}}.
\end{equation}
 The general solution of equation (\ref{loss}) is
\begin{equation}\label{solution}
E_{\widetilde{\tau}}=\left(E_{\widetilde{\tau}}^{'}+\epsilon\right)\exp(-x/\xi)-\epsilon.
\end{equation}
The left side of (\ref{solution}) is to be interpreted as the
residual energy of a $\widetilde{\tau}$ of initial energy
$E_{\widetilde{\tau}}^{'}$ after penetrating a depth $x$ of
material. \\

For those $\widetilde{\tau}$ ranging into the detector and fixing energy at
$E_{\widetilde{\tau}}$, they can be contributed by any initial
$\widetilde{\tau}$ with the energy above $E_{\widetilde{\tau}}$ and being
produced at the distance
$R(E_{\widetilde{\tau}}^{'},E_{\widetilde{\tau}})$, therefore,
 the differential flux intensity can be expressed:
\begin{equation}\label{spectrum}
\frac{\mathrm{d}N_{\widetilde{\tau}}}{\mathrm{d}E_{\widetilde{\tau}}}=
2\pi\int_{0}^{\frac{\pi}{2}} \sin\theta\mathrm{d}\theta
\int^{R}_{0}N_{A}\rho(r)\mathrm{d}x \int _{E_{\nu}^{th}}
\mathcal{I}_{\nu}(E_{\nu},x')\frac{\mathrm{d}\sigma_{\widetilde{\tau}}}
{\mathrm{d}E_{\nu}\mathrm{d}E_{\widetilde{\tau}}^{'}}\mathrm{d}E_{\nu},
\end{equation}
Here, we define the zenith angle $\theta $ as the angle between
the incident direction of neutrinos and the direction of the line
linking the center of both the earth and the detector. $N_{A}$ is
Avogadro's number, $\rho(r)$ corresponds to the earth density, and
$r=\sqrt{(x^{2}+R^{2}_{\oplus}+ 2xR_{\oplus}\cos\theta)}$ is the
distance from the center of the earth, where $R_{\oplus}$ refers
to the radius of the earth. $E_{\nu}^{th}$ is neutrino threshold
energy for $\widetilde{\tau}$ productions.
$\frac{\mathrm{d}\sigma_{\widetilde{\tau}}}
{\mathrm{d}E_{\nu}\mathrm{d}E_{\widetilde{\tau}}^{'}}$ refers to
the differential cross section of $\widetilde{\tau}$ productions.
$\mathcal{I}_{\nu}(E_{\nu},x')$ and $E_{\widetilde{\tau}}^{'}$ are
given by Equation (\ref{atte}) and (\ref{solution}), respectively.
We take
$x'=2R_{\oplus}\cos\theta - x$ instead of the distance a neutrino travelling in the earth.\\

\section{Possible signals analysis }
\mbox \\

High energy $\nu N$ processes can produce $\widetilde{\tau} $ as
well as muons. Obviously, the upgoing muons, as the background
flux, will range into the detector accompanying with  
$\widetilde{\tau}$ by similar interacting process.
 There are several possible ways exist to distinguish the
 $\widetilde{\tau}$ signals from the background muons.\\
\begin{figure}
\scalebox{0.5}{\includegraphics*[0pt,170pt][540pt,670pt]{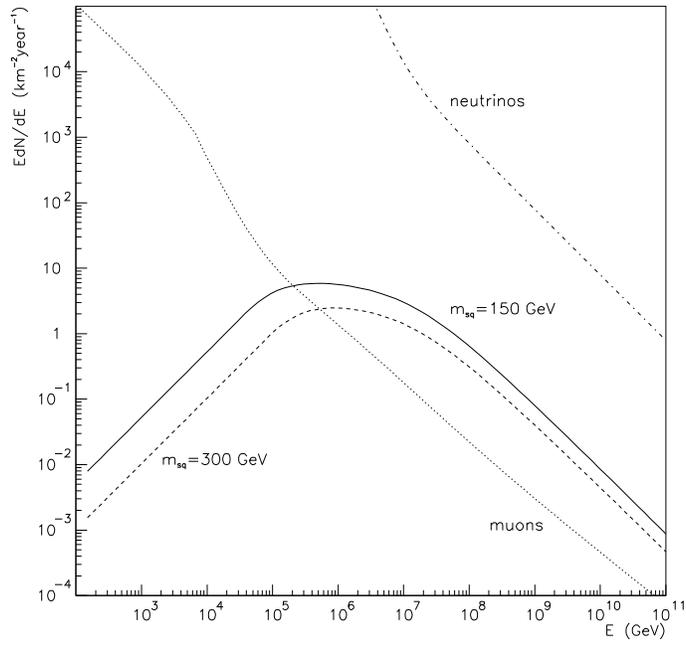}}
\caption{ \label{spectra} Energy spectrum of $\widetilde{\tau}$
pair events per $km^{2}$, per year, for $m_{\widetilde{q}}=150,
300 GeV$. Also shown are the upgoing neutrino and muon flux
through the detector. We make use of atmospheric neutrion flux with energy above
$1GeV$ adding the conservative extraterrestrial WB Limit as the
 incoming neutrino flux.}
\end{figure}

The first is its differential energy spectrum. Some detectors have
very well energy resolution, typically, L3+C detector\cite{l3c},
can reconstruct the exact tracks of charged particles in the
magnetic field, which can determine the momentum, charges and
direction of the incident particles at last. This can possibly
provides the direct evidences for some exotic particles.
 In Figure \ref{spectra} we show the energy
distribution for the $\widetilde{\tau}$ pair events for two choices of
$\widetilde{q}$
masses: $150 GeV, 300 GeV$. Also shown is the upgoing neutrino flux
 as well as the energy
distribution of upgoing muons'.  We see that, the dominant
contribution of $\widetilde{\tau}$
 comes from the energy zones between $10^{5}\sim
10^{7} GeV$. This mainly because most of $\widetilde{\tau}$
 are produced in
the earth with the energy above $10^{5} GeV$ and range into
detector with high energy as well. However, with the energy of
neutrinos growing, the range of $\widetilde{\tau}$
 with the energy above about
$10^{8} GeV$ turns out to be lager than the diameter of the earth,
therefore, all initial $\widetilde{\tau}$
 can totally arrive at the detector,
which makes the events rate curve almost parallel with that of
incoming neutrinos. Comparing with the curve of $\widetilde{\tau}$
 flux, we can
see, the muon flux's is much stronger  at the low
energy zones($<10^{5}GeV$), and turn out to be weaker at
the high energy zones($>10^{5}GeV$). \\

\begin{figure}
\scalebox{0.5}{\includegraphics*[0pt,170pt][540pt,670pt]{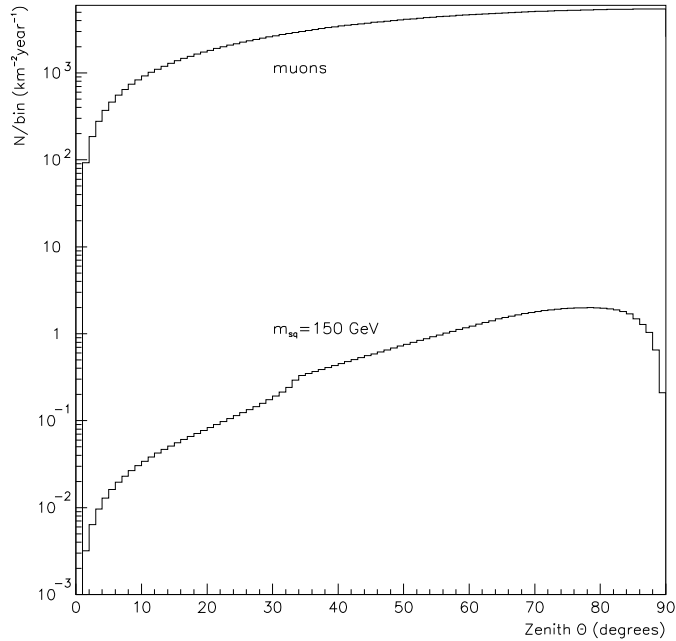}}
\caption{ \label{angle} Angular distribution of $\widetilde{\tau}$
pair events per $km^{2}$, per year, for $m_{\widetilde{q}}=150
GeV$. Also shown are the muon flux. We make use of atmospheric neutrion flux with energy above
$1GeV$ adding the conservative extraterrestrial WB Limit as the
 incoming neutrino flux. }
\end{figure}
The second is the angular distribution of events rate. Generally,
neutrino telescopes have poor energy resolution, only the angular
resolution is used to reduce the background. Figure \ref{angle}
shows us the angular distribution of events rate of 
$\widetilde{\tau}$ and muons.
 From the figure
one can conclude that the angular distribution of upgoing 
$\widetilde{\tau}$  is
definitely different from muon's. The dominant contribution of
upgoning $\widetilde{\tau}$ comes from the solid angle between $70\sim
80$ degrees. This is mainly because most of $\widetilde{\tau}$
 have their energy
between $10^{5}\sim 10^{7}GeV$ which hold the range about $10^{9}
cmwe$, this approximately equal to the acclivitous thickness of
the earth at the directions between $70\sim 80$ degrees.
Therefore, to the isotropic incoming neutrinos within that
directions, once interact with nucleons, the produced 
$\widetilde{\tau}$ will
range into the detector totally. For most of muons, their ranges
are neglectable comparing with the diameter of the earth, so the
events will grow as the angle increasing. It is a pity that one
 can hardly reduce background through this method duo to the
 peak signals of angular distribution completely overlayed by
 muons'. Comparing with extraterrestrial neutrinos, atmospheric
neutrino flux gives dominant contribution to the muon production.  \\

\begin{figure}
\scalebox{0.5}{\includegraphics*[0pt,170pt][540pt,670pt]{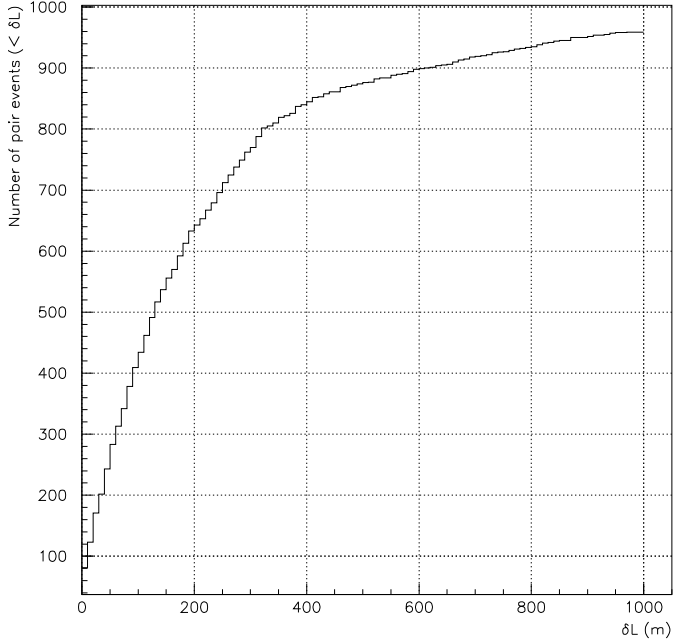}}
\caption{ \label{mc} The Monte Carlo result for integral total 
1000 pair events vs. their distance $\delta L$.
 We make use of atmospheric neutrion flux with energy above
$1GeV$ adding the conservative extraterrestrial WB Limit as the
 incoming neutrino flux. }
\end{figure}

The third is the tracks of the $\widetilde{\tau}$.
 The NLSPs are produced in
pairs and that will promptly decay into $\widetilde{\tau}$
 with the average
energy $(E_{\widetilde{\tau}}^{'})_{1,2}\approx 70\%(20\%)E_{\nu}
$ \cite{bwz}. As mentioned in Ref.\cite{albu},
typical signal events include two tracks separated
by certain distance
\begin{equation}\label{track}
\delta L \approx D \vartheta,
\end{equation}
where $D$ refers to the distance to the production point, and
$\vartheta$ is the angle between the pair particles. Considering
$D$ to be the same order with $\widetilde{\tau}$
 range, typically several
hundred kilometers and $\vartheta$ within $10^{-3}$, $ \delta L$
should typically be within several hundred meters. Figure \ref{mc}
gives a Monte Carlo result for 1000 $\widetilde{\tau}$
 pair events. We make use of atmospheric neutino flux adding extraterrestrial
neutrino flux(WB) as input and consider an exact interacting process
inside the earth, including different incoming directions,
attenuation of neutrinos and 
$\widetilde{\tau}$, enery conversion from a neutrino to two
 $\widetilde{\tau}$, the range of a  $\widetilde{\tau}$ running in the earth,
etc. At last we get an integral
distribution for number of pair events$(<\delta L, 10 m/bin)$ vs. the
distance $\delta L$,
 from which one can conclude that typical pair
events hold the $\delta L$ about $120m (\sim 50\%)$ and most of them are
within the confines of $320 m (\sim 80\%)$. Owing to all muon signals are
single tracks, seeking double tracks signals turns out to be an
effective method to distinguish from background. if we take
the two tracks which both enter into a detector within $\delta t=4$ microsecond 
as a double-track event, then for per $km^{2}$ per $year$, 
the accidental coincidence rate of muons is  as few as 
$ 2N_{\mu}N_{\mu}\delta t\approx 6\times 10^{-2} $, where 
$N_{\mu}$ refers to the number of events for muons.
\\

 In Table \ref{rate} we show the events rate for $\widetilde{\tau}$ pair
production per $km^{2}$ per year on the atmospheric neutrino flux with 
the energy above $1GeV$ adding 
different extraterrestrial neutrino fluxes. For
being comparable, we also show the rates of upgoing muons.
Comparing with the results from Ref.\cite{albu}, the number of events 
are increased at least an order of magnitude. This mainly because we work
under the SUSY breaking scenario with a quintessino LSP instead of a 
gravitino LSP. According to the analysis given above,  some large scaled  detectors
appear to be sensitive to the relatively long lived $\widetilde{\tau}$.

\begin{table}
\caption{\label{rate} Number of events per $km^{2}$ per year for taking
WB, AMANDA+max.extra-galactic p and MPR as extraterrestrial 
fluxes, respectively. Atmospheric neutrino flux with the energy above $1GeV$
 is taken into account as well.
The first column
refers to upgoing muons.  The
last two columns correspond to upgoing $\widetilde{\tau}$
 for two different
choices of squark masses: $150, 300 GeV$. }
\begin{center}
\begin{tabular}{|l||ccc|}\hline
&muon &$m_{\hat{q}}=150GeV$&$m_{\hat{q}}=300GeV$\\
 \hline
Atmos.+ WB&$4.83\times 10^{5}$&69&26\\
\hline
Atmos.+ AMANDA+max.&$4.90\times 10^{5}$&470&139\\
\hline
Atmos.+ MPR&$5.02\times 10^{5}$&3164&1244\\
\hline
\end{tabular}
\end{center}
\end{table}

\section{Conclusions}
\mbox \\

We discussed the exact production process of upgoing sleptons
based on the long lifetime scenario with a quintessino LSP.
 Furthermore, we detailedly
investigated the possible signals of events: energy spectrum,
angular distribution and their tracks, from which one can conclude
that the characteristic signals of events are distinctively
different from that of background muons'. The study shows that seeking
double-track upgoing events is a feasible method to detect 
$\widetilde{\tau}$ signals.
It's worth to mention that we introduced the 
contribution from atmospheric neutrino flux and found it's 
very important for upgoing $\widetilde{\tau}$ and muon
production. The event rates
are also given, the quintessino-LSP scenario predicts more NLSPs than 
that of gravitino-LSP. The numeric results show that km-scale detectors
are hopeful to get some positive results.\\

The possible signals may provide a direct evidence for
supersymmetry theory, furthermore, it can also offer a potential
solution for dark matter problem.

\section*{Acknowledgments}
\mbox \\

 We thank prof. LinKai Ding, QingQi Zhu and Dr ShuWang Cui for helpful
 discussions.
 This work is supported in part by the National Natural
  Science Foundation of China
  under the grant No.19999500.

\end{document}